\documentstyle[aps,twocolumn,floats,amsmath,graphicx]{revtex}

\begin{document}
\draft
%
%
\title{Measuring the Viscosity and Time Correlation 
Functions in a Microscopic Model of a Microemulsion}
\author{Songyun Qi and C.~M.~McCallum}
\address{Department of Chemistry, University of the Pacific, %
	3601 Pacific Ave., Stockton, CA  95211}
%
\date{April 22, 1999 (Submitted to \emph{Phys.~Rev.~E})}
\twocolumn[\hsize\textwidth\columnwidth\hsize\csname @twocolumnfalse\endcsname
\maketitle
\begin{abstract}
A dynamical lattice model is used to study the viscosity and the velocity-%
velocity autocorrelation function in a microemulsion phase.  We find 
evidence of anomalous viscosities in these phases (relative to water-rich
and/or oil-rich phases), in qualitative agreement with other results.
We also investigate the dynamic relaxation in the microemulsion phase.
It has been suggested that the temporal relaxation in
the microemulsion phase may be described by a stretched exponential 
Kolrausch-Williams-Watts law 
({\protect$\phi(t) =  A\exp[-(t/\tau)^{b}]$}).   In our model, we find
the velocity-velocity autocorrelation function $\langle v(0)v(t)\rangle$ 
fits this law, showing both enhanced ($b>1$) and inhibited ($b<1$) diffusion.
\end{abstract}
]
%

%
%
\section{Introduction}
\label{sec-intro}
 Using models to study the microemulsion phase has been made difficult 
 by the combination of short-ranged forces, short-range order, and 
 long-ranged disorder exhibited.  A large system is needed in order to 
 observe the unique isotropic ``structure'', while computer power 
 dictates what is actually feasible.  Very large systems can be 
 handled by Monte Carlo (MC) simulations (either continuous or 
 discretized), but the use of time-independent, non-physical moves 
 prevents calculation of temporal observables.  Molecular dynamics 
 (MD) allows the measurement of interesting time-dependent properties, 
 but also requires large numbers of independent particles.  The upper 
 capacity of MD is being continually 
 stretched~\cite{jb,neuvo98,tarek98,vanOs93,smit90}, but 
 very large bulk systems are rarely effectively modeled.

 There have been a number approaches have been used to balance these 
 concerns~\cite{gompper98,willemsen98,martins98,blake95,stassen94,%
 gerits92,boghosian97}.  In this paper, a dynamic lattice model 
 introduced earlier~\cite{jrg2} is further explored in which a 
 Boltzmann velocity field is mapped onto a nearest-neighbor lattice 
 model~\cite{zk,jrg1}, and the velocity field then is integrated into 
 the metropolis MC scheme.  It enables the determination of properties 
 usually only accessible by Molecular Dynamics methods, and has been 
 used to determine the diffusion constants and the dynamic surface 
 tension between competing phases, with evidence of a corresponding 
 wetting transition~\cite{cmm-jrg}.
\section{Dynamic Monte Carlo}\label{subsec-DMC}
One of the aims in investigating complex surfactant systems is to 
generate a method which can be used to study both equilibrium and 
dynamic properties.  Towards this goal, a general dynamic lattice MC 
model was constructed by combining Monte Carlo and Molecular Dynamic 
methods.  This so called Dynamic Monte Carlo (DMC) simulation 
\cite{jrg2} takes advantage of both methods, and is able to be 
utilized in both equilibrium and nonequilibrium studies.  This is used 
to investigate the dynamics of a lattice Monte Carlo, while keeping 
the equilibrium states of the model intact.

\subsection{Description of the DMC Model}\label{sec-dmc}
The DMC simulation algorithm incorporates pseudo velocities into Monte 
Carlo calculations within a three-dimensional lattice model.  In the 
base lattice model, cubic site $i$ is occupied by exactly one type of 
the three chemical species in a complex surfactant system.  These are 
denoted by a statistical variable $\sigma_{i}$, which can take on the 
values $\{-1,0,+1\}$, corresponding to cubes of oil, surfactant, or water 
respectively.  A vector variable $S_{i}$, defined at each lattice 
site, specifies the orientation of amphiphile particles.  The water 
and oil sites give no orientation.  This may be accomplished by the 
construction of a hybrid variable:
    \begin{equation}
        \label{eq:rho-hybrid}
        s_{i} = (1-\sigma^{2}_{i})S_{i}
    \end{equation}
Thus, only surfactant sites have a non-zero $s_{i}$.  The variable 
$S_{i}$ is normalized as unit vector quantity that varies continuously 
in three dimensions.

The principle advantage of lattice models, compared to continuum 
models, is that they dramatically simplify the simulated motions of 
the particles.  The particles may be located only at lattice sites, 
and there are no vacancies in the lattice.  Therefore, each site 
cannot exchange with other sites independently.  Only pairs of sites, 
together, can be considered for exchanges.  This exchange is achieved 
by considering both the relative velocity between two nearest-neighbor 
sites, and the potential energy through the Metropolis algorithm.  
Although the simulation does not reflect realistic molecular sizes or 
individual molecular movement, the simplicity and short interaction 
range make the lattice model far easier to study in detail than 
continuum models, and the qualitative results obtained from the model 
well represent the macroscopic behavior of oil-water-surfactant 
systems.

In order to depict dynamics with this equilibrium model, a velocity 
field is applied to the lattice MC. The temperature of the simulation 
box fixes the kinetic energy of the particles in the system (a form of 
equipartition):
\begin{equation}
    \frac{1}{2}\sum_{i=1}^{N} m_{i}v_{i}^{2} =%
    \frac{3}{2}N k_{B}T \quad .
    \label{eq:equi-full}
\end{equation}
At some time $t=0$ (really the first DMC step), the system is heated by 
randomly assigning velocities chosen from a Maxwell-Boltzmann 
distribution.  The Maxwell distribution in one dimension is a special 
case of the Gauss(ian) normal distribution.  Therefore, in each of the 
coordinate directions, the velocity components $v_{x}$, $v_{y}$, and 
$v_{z}$ obey the Gaussian distribution law.  These velocities must be 
identified as an average or center-of-mass velocity of all the 
molecules within the lattice site, because of the multi-particle 
nature of each site.  Thus, for a given temperature, we have
\begin{equation}
    \left\langle m_{i}v_{\alpha\,i}^{2}\right\rangle = k_{B} T \quad ,
    \label{eq:equi-one}
\end{equation}
where $\alpha$ is one of the three cartesian directions.

In the simulation, the movement of particles is achieved through the 
exchange of a pair of nearest-neighbors.  As stated earlier, the 
energy of the system is independent of the orientation of the water 
and oil sites.  Thus, including all desired features (treating water 
and oil sites as inversely symmetric, and treating the two ends of 
surfactant molecules symmetrically), the nearest-neighbor Hamiltonian 
may be written as
\begin{eqnarray}
    {\cal H} = \frac{1}{2}{\sum_{%
    \left\langle\mathbf{n},\mathbf{n'}\right\rangle}}' \bigg[ &%
    c_{1}\sigma_{\mathbf{n}}(\Delta\mathbf{n}\cdot %
    S_{\mathbf{n}'}) + %
    c_{2}\sigma_{\mathbf{n}}^{2}%
    (\Delta\mathbf{n}\cdot S_{\mathbf{n}'})^{2} & \nonumber \\%
    & + c_{3}\sigma_{\mathbf{n\phantom{'}}}\sigma_{\mathbf{n}'} +%
    c_{4}\sigma_{\mathbf{n\phantom{'}}}^{2}\sigma_{\mathbf{n}'}^{2}\bigg] &
    \nonumber \\[2ex]
    & + \sum_{\mathbf{n}}\big[\alpha\sigma_{\mathbf{n}}^{2} +%
    \omega\sigma_{\mathbf{n}}\big] &
    \label{eq:dmc-ham}
\end{eqnarray}
where $\Delta\mathbf{n}\equiv \mathbf{n} - \mathbf{n}'$ denotes the 
lattice difference vector between two sites $\mathbf{n}$ and 
$\mathbf{n}'$; $\alpha$ and $\omega$ are the effective chemical 
potentials of water and oil; and the coupling constants $c_{\alpha}$ 
determine the bonding energies ($c_{1}= 5/2$, $c_{2} = 3/4$, $c_{3} = 
-3/8$, and $c_{4} = -1/6$).  This creates a phase-diagram which 
includes a microemulsion phase.  This nearest-neighbor sum runs over 
pairs of sites; each site has $2d$ neighbors in $d$ dimensions.  By 
summing over all the particles in the system, the potential energy of 
the entire system is obtained.

At each simulation step, a pseudo-random walk is chosen through 
simulation space.  A switch variable is used to choose one direction 
($x$, $y$, or $z$), which ensures that each pair of nearest-neighbors 
are visited evenly.  For each trial exchange, the change of total 
energy $\Delta E$ is obtained by evaluating both potential and kinetic 
energy differences.  The potential energy factor is the difference of 
potential energy before and after the trial exchange.  The kinetic 
energy is measured by the relative velocity $v = v_{\mathbf{n}}- 
v_{\mathbf{n}'}$ only in the exchange direction.  This means only the 
component of the velocity in the exchange direction is considered, and 
the components in the other two directions are unchanged through the 
trial exchange (and therefore any accepted exchange).  The actual 
rules give results for the two possible signs of $v$: for the case $v 
< 0$, no exchange occurs even for a favorable $\Delta E$, as a 
negative relative velocity corresponds to two particles moving away 
from each other.  For $v>0$, a resulting (new) relative velocity $v'$ 
is obtained:
\begin{equation}
    \label{eq:vtovprime}
    \exp[-mv^{2}/2k_{B} T] = 1 - \exp[-m {v'}^{2}/2k_{B} T]
\end{equation}
In this manner, the final velocity components in the exchange 
direction are determined.

Both relative velocities and change of potential energy are considered 
for the change of total energy, and the acceptance or rejection of the 
exchange is then determined by the Metropolis algorithm.  This implies 
that the kinetic energy can supply potential energy required exchange, 
or there can be mixing between the kinetic and potential energy.  
Since the velocities are initially assigned from a Maxwell-Boltzmann 
distribution, and retained throughout the trajectory, the potential 
energy should also be contained within a Boltzmann distribution of all 
the configurations obtained.  For an accepted trial move, the chemical 
species within the two sites are exchanged, and the velocity 
components in the exchange direction will have new values resulting 
from Eq.~(\ref{eq:vtovprime}); for a rejected trial move, an elastic 
collision is applied, so that the two particles move away from each 
other with an exchanged velocity component in the direction of 
collision.

The continuous spin variable $s_{\mathbf{n}}$ needs special treatment.  
Because surfactant molecules may move to neighbor sites, the 
orientation of newly created surfactant site must come from an 
equilibrated distribution to maintain detailed balance in the system.  
This is accomplished by creating a ``ghost'' vector orientation at 
each lattice site during the relaxation procedure.  The ghost 
orientation at each site is also produced by Eq.~(\ref{eq:vtovprime}), 
and acceptance comes by including the potential energy through the 
Metropolis algorithm.  Therefore, the ghost orientations maintain 
equilibrium.  When a surfactant moves to its neighbor site, it takes 
the ghost orientation of the site.  System size is maintained by 
periodic boundary conditions.

\subsection{DMC Simulation in Practice}\label{subsec-simulation}
The microemulsion phase is studied by Dynamical Monte Carlo (DMC) 
simulation.  Most simulations were performed on systems of $24\times 
24\times 24$ lattice sites.  A test of validity of the DMC method is 
examining the velocity distribution over time (DMC steps).  The 
initial velocities of particles are generated according to a 
Maxwell-Boltzmann distribution.  At the end of the simulation, 
velocities maintain the Maxwell-Boltzmann distribution as predicted.  
Additionally, there is the requirement that the DMC model maintains 
the equilibrium states from the standard MC. Results for the energies 
of systems of identical compositions from the two models, lattice MC 
and DMC, were indistinguishable.

\begin{figure}[hbp]
    \centering
    \includegraphics[scale=0.57]{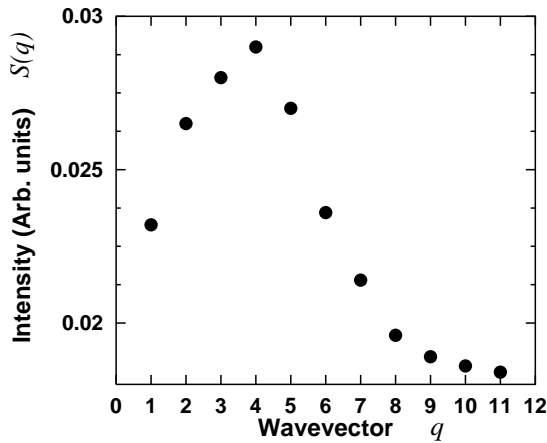}
    \caption{Water-water structure function of microemulsion phase 
    (T=1.5).}
    \label{fig:sk}
\end{figure}

Experiments indicate that the scattering intensity of microemulsion 
exhibits a peak at non-zero 
wavevector\cite{vd,shc,sb,gompper94,chen97a}.  This scattering 
behavior is considered the signature of microemulsion, in contrast to 
normal disordered phases (which show an exponential decay with 
increasing wavevector).  In the DMC, the scattering intensity is 
calculated as a discrete Fourier transformation of the water-water 
correlation function.  We have chosen as a ``typical'' microemulsion 
system: 26\% surfactant sites, 37\% oil sites, and 37\% water sites.  
An equilibrated microemulsion phase is obtained after 4000 to 6000 
simulation full lattice passes (DMC steps).  Simulated scattering data 
from such a phase is presented in Fig.~\ref{fig:sk}.

\section{Results and Discussion}\label{sec:results}
\subsection{Viscosity of Microemulsion}\label{sec:viscosity}
The investigation of the viscosity of surfactant systems is of great 
interest for both scientific and industrial applications.  Surfactant 
solutions show complicated rheological behavior.  It has been shown 
experimentally that the viscosity of such systems is affected by many 
factors such as concentration, molecular volume, presence of electric 
charges, hydrodynamic interactions, and temperature.  In practical 
terms, the viscosity of surfactant solutions is being scrutinized by 
industry.  For example, the flow rate is strongly affected when these 
solutions are pumped through pipes or stirred in large containers.  
The viscosity is also relevant to the design of surfactant 
solution-based flooding processes, which can be utilized in oil 
recovery (following the primary and secondary stages).  This is of 
great interest since some two-thirds of the oil originally in place 
can remain after the secondary recovery process.

The DMC has been used to investigate viscosity in microemulsion.  A 
(unit) velocity gradient was applied to the system in one direction of 
the 3D lattice, in order to simulate the behavior of fluids under 
shear stress.  The velocity component in $x$-direction ($v_{x}$) is 
forced to increase with increasing $y$.  Across each lattice spacing, 
$v_{x}$ is increased by a constant amount (the shear rate) in the 
$y$-direction.  The shear rate is a relative unit based on

 \begin{equation}
     \label{eq:shear-rate}
     \left\langle v_{\alpha}^{2}\right\rangle =%
     1 \quad ( \alpha = x, y, z) \quad .
\end{equation}

\begin{figure}[hbp]
     \centering
     \includegraphics[scale=0.40]{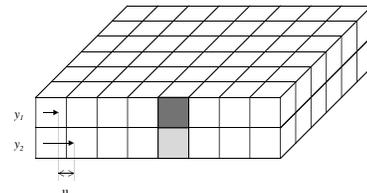}
     \caption{Schematic of the velocities of adjacent planes in the DMC.}
     \label{fig:exchange}
 \end{figure}

 Since all the lattice sites on the same $xz$-plane suffer the same 
 magnitude of shear, the exchanges between nearest-neighbors in the 
 $xz$-plane are allowed to be performed normally.  The exchange of 
 particles across the $y$-direction, which exchanges between two 
 adjacent $xz$-planes, needs to consider the different shear rates of 
 the two lattice sites.  The exchange of two particles by random 
 motion may happen between sites in planes $y_{1}$ and $y_{2}$ (see 
 Fig.~\ref{fig:exchange}).  
 Since  the particle in $y_{1}$ has a lesser velocity component in the 
 $x$-direction than does the particle in $y_{2}$, this exchange 
 process involves a net transfer of momentum in the direction of $y$.  
 For velocity components in the $y$- and $z$-direction ($v_{y}$ and 
 $v_{z}$), the values are simply exchanged without any added effect.  
 For the velocity component in the flow direction ($v_{x}$), the 
 relative velocity is calculated as before, but the net transfer of 
 momentum between the two planes is taken into account.  The particle 
 that moves from plane $y_{1}$ to $y_{2}$ is coming from a ``slower'' 
 velocity field compared to particles in plane $y_{2}$.  Therefore, 
 its local velocity should be lower than its neighbors.  On the other 
 hand, the particle from plane $y_{2}$ should have a higher velocity 
 than other particles in plane $y_{1}$.  This can be summarized by 
 the following expression for the final velocity component in the 
 $x$-direction:
\begin{subequations}
    \label{eq:vx}
    \begin{eqnarray}
        & v_{x}(y_{1}) = v_{x}(y_{2}) + \mathbf{u} & \\ 
	\label{eq:vxy1}
        & v_{x}(y_{2}) = v_{x}(y_{1}) - \mathbf{u} & \label{eq:vxy2}
    \end{eqnarray}
\end{subequations}
where $\mathbf{u}$ is the shear rate.

Viscosity has the generalized form\cite{dfe} \begin{equation} 
\label{eq:vis-Cvv} \eta = 
\frac{N}{T}\frac{1}{2t}\left\langle 
(C_{v\mathbf{u}}(t))^{2}\right\rangle
\end{equation} which uses the velocity-velocity correlation function 
$C_{vv}(t)$.  This can be re-written explicitly for $C_{vv}(t)$ with 
the result
\begin{equation}
    \label{eq:vis-vv}
    2t\eta = \frac{N}{T}\left\langle\left(\sum_{i=1}^{N}\left[%
     v_{ix}(t)\cdot\mathbf{u} - v_{ix}(0)\cdot\mathbf{u}%
     \right]\right)^{2}\right\rangle
\end{equation}

Use of the correlation function for viscosity measurements requires 
long lists of velocities for each particle at each DMC step.  This 
large physical memory requirement is quite restrictive.  Another way 
is to take advantage of the relative velocities used in the exchanges.  
Labeling this relative exchange velocity as $v_{ij,\alpha}$, we can 
use the sum
\begin{equation}
    \eta = -\sum_{i,j}\frac{mv_{ij,x}-\mathbf{u}}{\mathbf{u}L^{3}\tau}
    \label{eq:viscosity}
\end{equation}
This sum is over only accepted exchanges of two nearest-neighbors in 
the $y$-direction.  $L$ is the linear lattice size (generally, $L=24$ 
in this work).  The variable $\tau$ is introduced as the reduced 
time step, which ensures that the average velocity is equal to the 
average rate of exchange of the sites ($\left\langle |v|%
\right\rangle\tau = \left\langle P(v)\right\rangle$).  It is a 
constant of the simulation method.  One of the advantages of this 
method is that it can work with periodic boundary conditions, since 
the shear rate of one site is only relative to its neighbors, and can 
only be considered during the exchange.  Furthermore, since the flow 
is not explicitly introduced by increasing the velocity of each 
lattice site, and the kinetic energy is defined through local 
velocities (which are balanced during each exchange), any heating 
affect of the process is minimized.

Results for these viscosity measurements are presented in 
Table~\ref{tbl:vis-res}.  These results are for an equilibrated 
microemulsion phase (as determined by $S(k)$ data), at a constant 
temperature.  The temperature can be independently determined through 
the kinetic energy.  One can see that the apparent temperature falls 
as the shear rate is raised.  This is due to the increased 
transference of kinetic energy to potential energy, also reflected in 
a corresponding increase in the potential energy.
\begin{table}[h]
    \centering
    \caption{The viscosity from DMC of the microemulsion phase at 
    $T=1.0$.}
    \begin{tabular}{|c|c|c|c|}
        \hline
        Shear Rate & Shear Viscosity ($\eta$) & Internal Energy & 
        Final $T$  \\
         & $((\mathrm{m}J)^{1/3}/\ell)$ & $(J)$ & $(J/k_{\mathrm{B}})$  \\
        \hline
        0.005 & 2.282 & -1.448 & 0.979  \\
        0.010 & 1.862 & -1.450 & 0.983  \\
        0.015 & 1.523 & -1.461 & 0.983  \\
        0.020 & 1.371 & -1.465 & 0.974  \\
        0.025 & 1.305 & -1.469 & 0.969  \\
        0.030 & 1.227 & -1.474 & 0.965  \\ \hline
    \end{tabular}
    \label{tbl:vis-res}
\end{table}

\begin{figure}[hbp]
    \centering
    \includegraphics[scale=0.57]{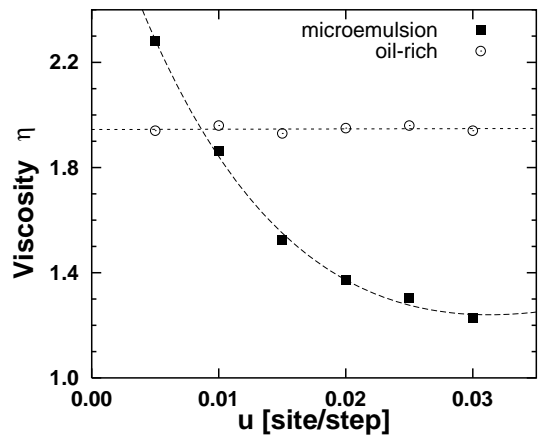}
    \caption{The dependence of viscosity measurements on shear rate.}
    \label{fig:vis-v}
\end{figure}

DMC calculations of the shear viscosity of microemulsion with shear 
rate changes exhibit some notable features, illustrated in 
Fig.~\ref{fig:vis-v}.
This bears some discussion about what is expected for such fluids.  
Currently, several studies of viscosity of microemulsion have been 
published in the literature.  Gogarty and Dreher\cite{kdd} have 
reported experimental results of microemulsion which shows strong 
non-Newtonian behavior.  Many researchers have given values of 
microemulsion viscosity at only a single value of shear rate, or shear 
stress \cite{da,rnh1,rnh2,scj,acosta98}.  In some cases, non-Newtonian 
or anomalous behaviors were observed.\cite{da,rnh1,rnh2,acosta98} 
However, there are several other researches which report Newtonian 
viscosities \cite{md,gbt,amb}.  This may result from the contradictory 
elements of small interaction range and weak velocity variation.  The 
hydrodynamic velocity gradients applied in these experiments are not 
high enough to create a strong velocity variation.  Thus the system 
needs a relatively longer time to allow the velocity variation to be 
transferred compared to the renewal time of the system structure.  
This limitation is not always apparent in the microemulsion phase 
containing comparable amounts of water and oil.  In particular, in a 
viscosity measurement of microemulsion obtained by a cone-and-plate 
rheometer, both Newtonian and non-Newtonian behaviors have been 
observed \cite{bmk}.

The data obtained from the DMC simulation exhibit a strong 
non-Newtonian behavior, which corresponds to only some of the 
experimental results.  There is still much work to be done in this 
area before one can draw any conclusions as to which experimental 
results are correct (perhaps both sets are).  However, we do know this 
behavior is not an artifact of the DMC, as water-rich (oil-rich) 
phases exhibit only Newtonian viscosities.  On the other hand, there 
are some shear-induced perturbations in the simulation which may 
affect the viscosity results.  The change of velocity component 
resulting from the shear is not an actual Boltzmann factor.  This 
should affect the results of Metropolis algorithm in determining the 
acceptance or rejection of the exchange in the following simulation 
steps.  In other words, a change of configuration is induced in the 
system corresponding to the shear rate.  Evidence of this is the 
change of the internal energy and temperature of the system which are 
observed at high shear rate.  Compared to the change of viscosity with 
shear rate, however, the affect of shear-induced error does not seem 
considerable.  At lower shear rates, the results are seemly linear.  
In any case, in order to minimize the perturbation of shear on the 
system, and statistical error in calculation, the viscosity is 
determined at a relatively low shear rate.  Thus, we shall now fix the 
shear rate at 0.01 (relative velocity units).

\begin{figure}[hbp]
    \centering
    \includegraphics[scale=0.57]{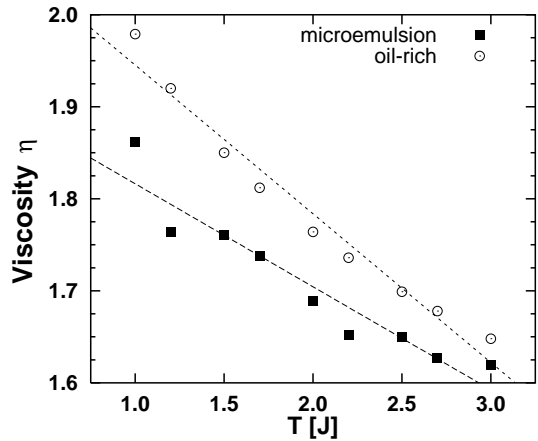}
    \caption{The dependence of viscosity measurements on temperature.}
    \label{fig:vis-T}
\end{figure}

Purely thermal effects are also interesting to consider.  
Heuristically, viscosity should decrease with increasing temperature, 
but the overall behavior of microemulsion viscosity with temperature 
is not well understood.  To investigate phase behavior of $\eta$, the 
shear viscosity of microemulsion phase was calculated and compared 
with the oil-rich phase at different temperatures.  These results are 
presented in Table~\ref{tbl:vis-comp} and Figure~\ref{fig:vis-T}.
\begin{table}[ht]
    \centering
    \caption{Shear viscosity of the microemulsion phase 
    and the oil-rich phase ($\mathbf{u} = 0.01$).}
    \begin{tabular}{|c|c|c|c|}
        \hline
        Temperature & $\eta$ (microemulsion) & $\eta$ (oil)  & \\
         & ($({\mathrm m}J)^{2}/\ell$) & ($({\mathrm m}J)^{1/3}/\ell$)  
         & \% Change \\
        \hline
        1.0 & 1.862 & 1.979  & 5.9\\
        1.2 & 1.764 & 1.920  & 8.1\\
        1.5 & 1.761 & 1.850  & 4.8 \\
        1.7 & 1.738 & 1.812  & 4.1 \\
        2.0 & 1.689 & 1.764  & 4.2 \\
        2.2 & 1.652 & 1.736  & 4.8 \\
        2.5 & 1.650 & 1.699  & 2.9 \\
        2.7 & 1.627 & 1.678  & 3.0 \\
        3.0 & 1.619 & 1.648  & 1.8 \\
        \hline
    \end{tabular}
    \label{tbl:vis-comp}
\end{table}

One can see that with an increase in temperature, the viscosity of 
both phases decrease, which is characteristic of all liquids.  At a 
given temperature, the microemulsion phase shows a considerably lower 
viscosity than the oil-rich phase, and this difference decreases with 
increasing temperature.  Such behavior of microemulsion corresponds to 
the general experimental fact that the microemulsion phase exhibits 
low viscosity compared to other bulk phases.

It has been found experimentally that the viscosity of microemulsion 
is strongly affected by its microstructure.  Papaioannou and 
coworkers\cite{sis6} have published viscosity measurements of 
microemulsion as a function of salinity.  They found that the viscosity 
of microemulsion exhibits two peaks: one at the oil-rich, and one at 
the water-rich phase boundaries.  Additionally, a minimum was measured 
in the Winsor III state.  It was also found through other experiments 
that the viscosity of microemulsion decreases as the structure becomes 
more and more bicontinuous.\cite{sis10} In the DMC simulation, the 
microemulsion phase has comparable amounts of oil and water, 
corresponding to the bicontinuous Winsor III state.  The DMC therefore 
shows good agreement with the experimental results.

\subsection{Long-time Behavior of the Velocity Auto-correlation Function}
\label{sec:long}
Use of the time-correlation function is one of the important 
approaches for describing transport processes and other time-dependent 
phenomena.  This general function is considered to play a similar role 
in non-equilibrium Statistical Mechanics as the partition function 
plays in equilibrium situations.  One advantage of the 
time-correlation function is that the resulting transport coefficients 
are quite general and independent of the details of any particular 
model.  In particular, the self-diffusion coefficient can be expressed 
in terms of the velocity auto-correlation function as
\begin{equation}
    D = \frac{1}{3}\int_0^\infty \left\langle v(0)\cdot 
    v(t)\right\rangle\,\mathrm{d}t \quad .
    \label{eq:diff}
\end{equation}

This expression is valid for any classical system in which the 
diffusion is governed by the diffusion equation (Fick's law).  
Therefore, the velocity auto-correlation function is a valuable 
approach to investigate the diffusion behavior of complex systems.  
Autocorrelation expressions like Eq.~(\ref{eq:diff}) may be nicely 
related to experiment.  Observation of relaxation is a common 
experimental method to investigate the structure and dynamics of 
systems.  Because both equilibrium and dynamical properties of the 
symmetric microemulsion phase are still very uncertain, the relaxation 
are of continuing importance.

The general form of velocity auto-correlation function is
\begin{equation}
    C_{vv}(t) = \left\langle v(0)\cdot v(t)\right\rangle \quad .
    \label{eq:acf}
\end{equation}
When $t = 0$, $C_{vv}(t)$ simply has the value of equilibrium average 
of $v^{2}(0)$ ($3k_{\mathrm{B}}T/m$ by equipartition).  As time 
advances, $v(t)$ will be less
correlated with its initial value.  Thus, the velocity 
auto-correlation function is expected to decay to zero from its 
initial value.  In general, it is assumed that any $C(t)$ decays 
exponentially with a time constant $\tau$.  This can be expressed as
\begin{equation}
    C(t) = A\exp\{-t/\tau\} \quad .
    \label{eq:C-exp}
\end{equation}

However, it has been found experimentally that time-relaxation in many 
disordered systems does not follow the simple exponential decay.  The 
microemulsion phase is one example of this.  Losada and 
L\'opez-Quintela~\cite{mal} found a relaxation mode in microemulsion 
which follows a Kohlrausch-Williams-Watts (KWW) stretched-exponential 
law.  The general form of this law is
\begin{equation}
    \label{eq:kww}
    \Phi(t) = A\exp\{-(t/\tau)^{b}\} \quad , \quad (0\leq b \leq 1)
\end{equation}
The super-exponential factor $b$ was found to be dependent on many 
conditions, such as the spatial, temporal and energetic disorder of 
the system.  There has been only limited discussion of the 
applicability of this law, and its 
ramifications~\cite{lopezq91,weiss94,vanderhoef90}.

In order to measure the velocity auto-correlation function in the DMC 
model, the derivative of the mean-squared displacement is used:
%
%
\begin{equation}
    \label{eq:acf-deriv}
    \frac{\partial \left\langle [r(t) - r(0)]^{2}\right\rangle}%
    {\partial t} =\left\langle v(t)\cdot v(0)\right\rangle \quad .
\end{equation}
In the DMC method, since the movement of particles is achieved only by 
the exchange of a pair of nearest-neighbors and is not explicitly 
decided by velocity of each single lattice site, a ``measured'' $\left\langle 
v(t)\cdot v(0)\right\rangle$ will not reflect the actual diffusive 
behavior of the system.  The velocity auto-correlation function is 
therefore defined through Eq.~(\ref{eq:acf-deriv}).

\begin{figure}[h]
    \centering
    \includegraphics[scale=0.5]{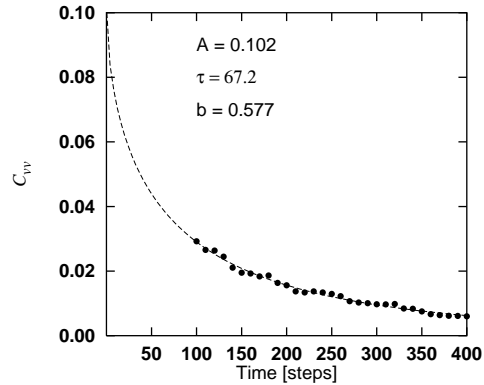}
    \caption{ KWW stretched-exponential law fit to data of velocity 
auto-correlation function of microemulsion phase (T=1.5).}
    \label{fig:kww}
\end{figure}

Because of the average assembling nature of the 
velocity auto-correlation function, the long-time behavior is more 
difficult to observe and less accurate \cite{jev}.  In determining 
whether relaxation follows a KWW-type law, the longest-time 
correlations are the most important.  This requires extremely long 
simulation (cpu) times and proper equilibration before results may be 
examined.  For a typical simulation, he 
velocity auto-correlation function is obtained from 3600 to 4000 
equilibrated simulation lattice passes.  Data obtained at different 
temperatures were fit to Eq.~(\ref{eq:kww}) for the longest part
(last $2/3$) of simulation time --- 1000 to 4000 DMC 
steps (Fig.~\ref{fig:kww}).  The fit is satisfactory for all 
microemulsion data obtained, and better than a simple exponential fit 
(Eq.~(\ref{eq:C-exp})).  Although \textbf{all} data (from $t=0$) were 
fit satisfactorially with either Eq.~(\ref{eq:C-exp}) or 
Eq.~(\ref{eq:kww}) (with $b\approx 1$), when \textbf{only} 
the longest-time data were included, the simple exponential fit was 
inadequate.  Thus, Eq.~(\ref{eq:kww}) fits the entire data range 
well, while Eq.~(\ref{eq:C-exp}) breaks down at longer ``times''.  
Additionally, auto-correlation data obtained 
from the DMC in the water-rich or oil-rich phases can not be found to 
fit a ``pure'' ($b\neq 1$) KWW law.  The values of the KWW fit
parameters are presented in Table~\ref{tbl:kww-dep}.

\begin{table}[h]
    \centering
    \caption{Temperature dependence of the KWW equation parameters in 
the microemulsion phase (A=0.1).}
    \begin{tabular}{|c|c|c|}
        \hline
        Temperature & $\tau$ & $b$ \\
        \hline
        2.5 & 145.63 & 1.365 \\
        2.2 & 121.67 & 1.054 \\
        2.0 & 105.79 & 0.840 \\
        1.7 & 82.48 & 0.630 \\
        1.5 & 69.67 & 0.575 \\
        1.3 & 58.55 & 0.490 \\
        1.2 & 53.65 & 0.470 \\
        1.1 & 50.48 & 0.459 \\
        1.0 & 48.71 & 0.418 \\
        \hline
    \end{tabular} 
    \label{tbl:kww-dep}
\end{table}

Both the stretched exponent $b$ and the macroscopic relaxation time 
$\tau$ are found to increase with increasing temperature within the 
microemulsion phase.  The increase in relaxation time may be 
understood as a slowing of the diffusion rate as the system approaches the 
critical temperature.  A number of reports have been published 
on the experimental study of critical phenomena of surfactant 
systems~\cite{sis10}.  However, the critical behavior in surfactant 
solutions is still far from being understood completely.  One common 
feature found in most of the investigations is that phase separation 
occurs as temperature is raised.  This may be related to the slowing 
of diffusive behavior, observed in the present model.

The wide variation of $b$ is of great interest.  Notably, a value of 
$b$ greater than 1 was obtained at high temperatures, which has not 
been observed in normal disordered phases.  This corresponds to 
``enhanced diffusion'' --- faster than expected.  Losada~\cite{mal} 
has reported a relaxation study of microemulsion using a pressure-jump 
technique.  In this experiment, the relaxation behavior with $b>1$ was 
observed in a microemulsion phase that had similar concentrations of 
oil and water.  This has been the only experimentally-observed example 
of enhanced diffusion in microemulsion.  It is hoped that additional 
work in this area will be done to confirm this interesting and 
important result.

\begin{figure}[hbp]
    \centering
    \includegraphics[scale=0.57]{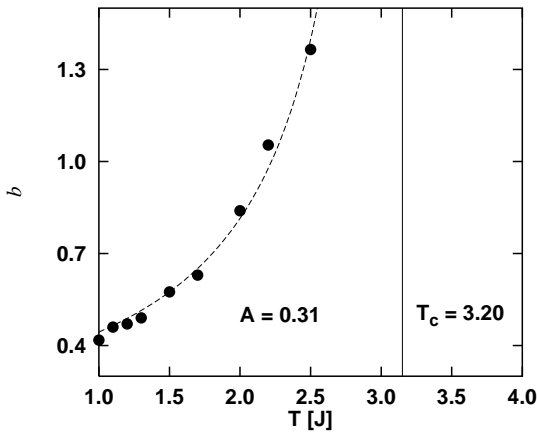}
    \caption{Temperature dependence of the stretched exponent $b$.}
    \label{fig:KWW-b}
\end{figure}

The stretched exponent $b$ is strongly temperature-dependent.  It has 
been treated by the theory of critical phenomena, and Losada found 
that it fit the critical equation
\begin{equation}
    b = A \, [(T_{c}-T)/T_{c}]^{-1}
    \label{eq:b-critical}
\end{equation}
The temperature dependence of $b$ was measured by DMC simulation in 
order to confirm this behavior.  The DMC data was fit to 
Eq.~\ref{eq:b-critical} (Fig.~\ref{fig:KWW-b}).  This fit resulted in 
the value $A=0.30$, and gave a critical temperature $T_{c} = 3.20$.  
From Fig~\ref{fig:KWW-b} one can see that $b=1$ when $T=2.21$; at this 
point, the relaxation (as well as the diffusion) is ``normal'' and the 
KWW equation reduces to the exponential form.  At temperatures below 
this point, the microemulsion shows an inhibited diffusive behavior.  
This can be related to the spatial and temporal disorder present in 
the system.  It has been found in both experimental and theoretical 
research that diffusion behavior moves from anomalous to normal as the 
temperature is increased\cite{wpk,jk}.  This corresponds to DMC 
results obtained for $T < 2.21$.  As the temperature is increased past 
this point, there is an enhancement of diffusion in the system.

%

\section{Conclusions}\label{sec:conclusion}
The complex systems formed between oil, water and surfactant molecules 
were examined with a three-dimensional lattice model enhanced by the 
addition of a coarse-grained velocity field.  The shear viscosity of 
microemulsion phase was measured as a function of shear rate, and 
showed a decrease as the shear rate increased.  The measurements were 
also performed on oil-rich phases, which showed typical behavior of a 
Newtonian (classical) fluid.  Compared to the oil-rich phase, the 
viscosity of microemulsion shows strong non-Newtonian behavior.  This 
corresponds to at least one experimental observation of the existence 
of non-Newtonian shear behavior in microemulsion~\cite{bmk,acosta98} The shear 
viscosities were also measured as a function of temperature.  In all 
the systems simulated here, the microemulsion phase shows a notably 
lower viscosity than the oil-rich phase.  As the temperature 
approaches to the critical temperature, the difference decreases.  We 
would expect the two viscosities to equal at the thermodynamic limit.  
These viscosity measurements in the microemulsion phase qualitatively 
reproduce experimental observations~\cite{sis6,sis10}.

Measurements of the velocity autocorrelation function were obtained by 
the distance displacement method.  The most interesting observation is 
that the long-time tails of microemulsion do not obey a simple 
exponential decay, but rather a Kohlrausch-Williams-Watts 
stretched-exponential law.  The same behavior has been observed 
experimentally in a few other disordered phases~\cite{mal,daoud90}.  
Microemulsion, a particular disordered phase, shows many properties 
distinct from normal disordered phases, which makes the long-time 
behavior very compelling.  Both the stretched exponent and the 
macroscopic relaxation time were found to be strongly 
temperature-dependent.  The temperature dependence of the 
super-exponential $b$, shows critical scaling.  The values of $b$ can 
be divided into three temporal ranges \mbox{$b=1, b<1, b>1$}.  Since 
the velocity autocorrelation function is directly related to the 
diffusion constant, this suggests the existence of normal, inhibited 
and enhanced diffusion in this model.  These results provide evidence 
that DMC method employed in this work should be capable of simulate 
not only the equilibrium but the dynamics of the system as well.  The 
diffusion constants and the dynamic surface tension between 
microemulsion and oil-rich phases has already been 
calculated~\cite{cmm-jrg}, and the measurement of other properties 
should also be possible through use of the DMC model.

Due to the fact that the molecules are defined at the lattice sites, 
the movements of particles are restricted to unit length.  
Consequently, the velocities involved in the calculations are not 
completely realistic.  In particular, a small spatial region of the 
velocity autocorrelation function is anti-correlated and inaccurate at 
certain degree.  Furthermore, we make no attempt to make quantitative 
comparisons of shear viscosity with experimental data.  Such 
comparison would require more realistic modeling of the molecular 
interactions.

One aspect of future work may connect velocity autocorrelation 
function results to the cooperative relaxation behavior of the system.  
The dipole correlation function is one way to describe of the 
mechanism of cooperative relaxation through experiment, so that the 
velocity autocorrelation function can be related to the relaxation of 
the dipole moment in droplets, as well as the structural and kinetic 
properties of the system.  The observed non-Newtonian behavior is a 
macroscopic representation of the microstructure of microemulsion.  
However, at present there is little theory to connect these viscosity 
observations with microstructural evolution.  Since both Newtonian and 
non-Newtonian regimes have been observed experimentally, this phase 
progression of microemulsion is of interest to explore further.
%

%
%

%
%
%

%
%
%
\end{document}